# Min-Time Escape of a Dubins Car from a Polygon

Isaac E. Weintraub[1], Alexander Von Moll[1], David Casbeer[1],
Satyanarayana G Manyam[2], Meir Pachter[3], and Colin Taylor[4]

*Abstract*—A turn constrained vehicle is initially located inside a polygon region and desires to escape in minimum time. First, the method of characteristics is used to describe the time-optimal strategies for reaching a line of infinite length. Next, the approach is extended to polygons constructed of a series of line segments. Using this construction technique, the min-time path to reach each edge is obtained; the resulting minimum of the set of optimal trajectories is then selected for escaping the polygon.

## I. INTRODUCTION

Path planning is an essential task for mobile automated systems. In this task, the objective is to obtain the trajectory or *path* that a vehicle should take to accomplish some objective subject to a set of constraints. In this work, a mobile vehicle that has a minimum turn radius is considered. The objective is for this vehicle to escape a convex polygon in minimum time, provided that the vehicle start inside the region at initial time. In light of this objective and these constraints, a number of related works should be highlighted.

Lester Dubins, known for his seminal works on vehicle path of minimum time stated that min-time paths for turn constrained vehicles resolved to a sequence of hard turn and straight line paths [1], [2]. Of course, this model assumed that the vehicle not reverse directions - this was later considered in great detail by Reeds and Shepp [3]. Dubins' analysis predates Pontryagin's Maximum Principle (PMP) [4], which (in short) states that the optimal control is one that optimizes the Hamiltonian of a dynamical system. Boissonnat, Cérézo, and

*This paper is based on work performed at the Air Force Research Laboratory (AFRL) *Control Science Center*. DISTRIBUTION STATEMENT A. Approved for public release. Distribution is unlimited; AFRL-2024-5262. This work is funded in-part by AFOSR, LRIR 24RQ-COR002.

[1]Weintraub, Von Moll, and Casbeer are with the Air Force Research Laboratory, 2210 8th St. WPAFB, OH 45433 `{isaac.weintraub.1, alerander.von_moll, david.casbeer}@afrl.af.mil`

[2]Manyam is with DCS Corp., 4027 Colonel Glenn Hwy, Dayton, OH 45431

[3]Pachter is with the Department of Electrical Engineering, Air Force Institute of Technology, WPAFB, OH 45433

[4]Taylor is with Parallax Research, 4035 Colonel Glenn Hwy. Beavercreek, OH 45431

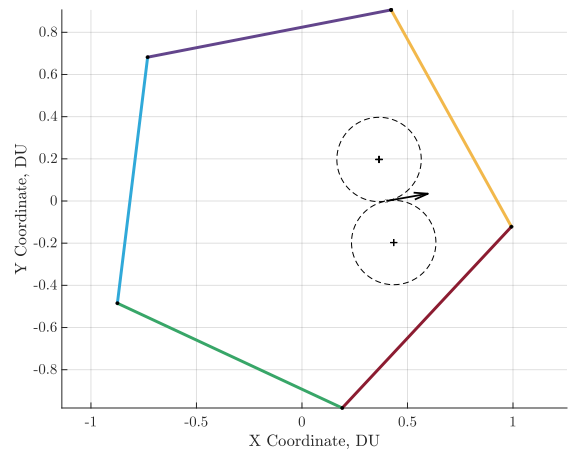

Fig. 1. A turn-constrained vehicle with constant speeds aims to escape a convex polygon in minimum time. The min-radius turn circles are drawn as dashed circles. The arrow represents the current location and direction of the vehicle, contained within the polygon.

Leblond showed that the resulting optimal control for a turn constrained vehicle exhibits "bang-zero-bang" phenomenon, consisting of hard-turn and straight-line paths for min-time strategies of turn-constrained vehicles [5]. A classification of the turn directions and classification of Dubins paths was performed by Shkel and Lumelsky in [6]; thereby increasing the computational efficiency of obtaining the minimum-time path of viable solutions.

More applications and generalization of Dubins paths have been identified and studied including path planning amidst obstacles, currents/drifts, and routes through multiple points. Path planning of a Dubins vehicle amidst obstacles consisting of line segments and circular arcs was analyzed in [7]; consideration of a reverse gear amidst obstacles can be found in [8]. Vehicle paths considering disturbances such as fluid currents were considered in [9]–[14]. Strategies for the min-time visiting of a Dubins vehicle visiting three points was obtained in [15]. Cases where many points were considered gave rise to the Dubins Travelling Salesman Problem (TSP) [16], [17] or a set of regions [18] or cliques [19].

This work considers the escape of a Dubins vehicle from a polygon region. Related to this premise, escape from a cir-cular target has been considered in [20]. The authors of this work; incrementally improved upon [20] by reducing the state space by one dimension in [21]. The method of characteristics [22] and optimal control [23], [24] is applied to the problem of min-time escape of a turn-constrained vehicle to an infinite line. These results motivate the utility of geometric tools that are then extended to the escape of a polygon region.

Other related works that include Dubins paths involving line-segments. In [25], the state space is partitioned to assist path planning through a field of obstacles. A set of lines called *Dubins Gates* were constructed and path plans to reach those lines were obtained in the course of that work. A TSP problem of visiting polygon regions is found in [26]. In their work, the objective was to reach each polygon region in minimum time. Lastly, the Dubins iterval problem of departing and reaching a point with constraints on the departure and arrival headings was investigated in [27]. This approach proves useful when the heading of the vehicle is required to not only reach a desired point in space but within a desired interval of headings.

The contributions of this work are as follows:
1) Time-optimal paths for a Dubins vehicle for reaching an infinite line are solved in closed form.
2) The Dispersal Line, Universal Line, Usable Part, Non-Usable Part, and Boundary of the Usable Part are identified in the infinite-line case.
3) The algorithm for extending the result from an infinite line to a convex polygon is shown.
4) The existence of the Dispersal Surface, where two time-optimal trajectories are present is described and highlighted in an example.

This work is constructed as follows: First the min-time trajectory is posed and solved in Section II; the dispersal line is described in Section II.A, the turn-only strategy is described in Section II.B, and the turn-straight strategy is described in Section II.C. Leveraging the solution by way of the method of characteristics and a geometric solution is described in Section III. In Section IV, the solution for the infinite line is extended to a polygon with any number of edges; an algorithm that shows how this can be performed is also described. Lastly in Section V, concluding remarks and future work is described.

## II. MIN-TIME TRAJECTORY TO A LINE

The dynamics are of the form: $\dot{\mathbf{x}}(t) = \mathbf{f}(\mathbf{x}(t), u(t), t)$, where the state space $\mathbf{x} = (x, \theta)^\top \in R \cup (-\pi, \pi]$ and the control $u(t) \in [-1, 1] \subset \mathbb{R}$. The state variables are composed of the distance of the vehicle from the line, having and the domain $x \leq 0$ and the heading relative to the $x$-axis, $-\pi < \theta \leq \pi$. The dynamics are as follows.

$$\dot{x}(t) = v \cos \theta(t),$$
$$\dot{\theta}(t) = \frac{v}{R} u(t), \quad -1 \leq u(t) \leq 1, \quad 0 \leq t \quad (1)$$

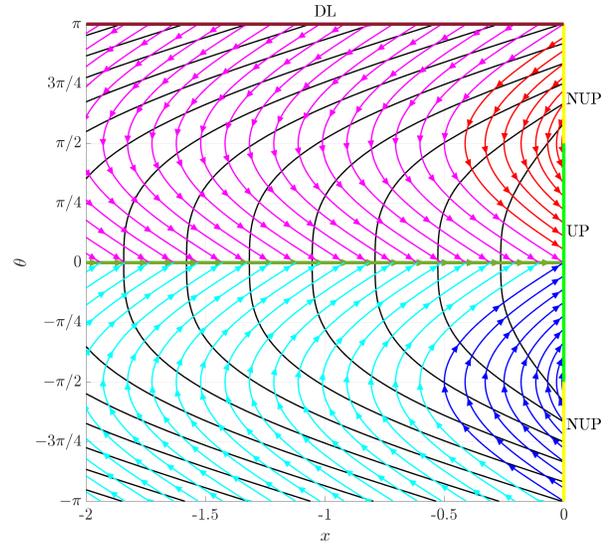

Fig. 2. Optimal flowfield for the min-time escape to a line

Define the terminal manifold to be a line located at the origin and spanning in infinite direction along the $y$-axis. The terminal manifold, $\mathcal{M}$ is as follows.

$$\mathcal{M} = \{(x, \theta) \mid x = 0, -\pi < \theta \leq \pi\} \quad (2)$$

Using the method of characteristics, one constructs the Usable Part (UP) of the terminal manifold as defined in (2).

$$\text{UP}(\mathcal{M}) = \left\{ (x, \theta) \,\middle|\, \min_{-1 \leq u(t) \leq 1} (\vec{n} \cdot \mathbf{f}) < 0 \right\} \quad (3)$$

Once the UP is obtained, equations of motion are then backward propagated in *retrograde form* from the UP and fill the state space with optimal trajectories. To construct the UP, $\vec{n}$ are the outward pointing normals from the terminal manifold into the state space, $\mathbf{f}$ are the dynamics, and $u(t)$ is the admissible control. The manifold is a line where $x = 0$ and the objective is to penetrate the line from the left ($x_0 < 0$). Therefore the outward pointing normals are $\vec{n} = (-1, 0)^\top$. Substituting the normals and the dynamics from (1), the UP from (3) is evaluated and is

$$\text{UP}(\mathcal{M}) = \left\{ (x, \theta) \,\middle|\, \min_{-1 \leq u(t) \leq 1} (-v \cos \theta) < 0 \right\}$$
$$= \left\{ (x, \theta) \,\middle|\, x = 0, -\frac{\pi}{2} < \theta < \frac{\pi}{2} \right\} \quad (4)$$

The optimal control needs to be formed and this enables one to construct the time-optimal trajectories for reaching the terminal manifold (the line) in minimum time. The objective cost functional is

$$u^*(t) = \operatorname*{argmin}_{-1 \leq u(t) \leq 1} \{J\} = \operatorname*{argmin}_{-1 \leq u(t) \leq 1} \left\{ \int_0^{t_f} 1 \, dt \right\} \quad (5)$$

To obtain the optimal control that minimizes the time for the vehicle to reach the terminal manifold, one begins by formulating the Hamiltonian, $\mathcal{H}$. The Hamiltonian is the running cost added to the inner-product of the costates and dynamics.

$$\mathcal{H} = 1 + \lambda_x v \cos\theta + \lambda_\theta \frac{v}{R} u \quad (6)$$

Taking partials of the Hamiltonian with respect to the states, the costate dynamics are obtained.

$$\dot\lambda_x = -\frac{\partial \mathcal{H}}{\partial x} = 0 \quad \dot\lambda_\theta = -\frac{\partial \mathcal{H}}{\partial \theta} = \lambda_x v \sin\theta \quad (7)$$

As shown in (7), $\lambda_x$ is constant for all time. From Pontryagin's Minimum the Hamiltonian resulting from the optimal control is a minimum.

$$\mathcal{H}^*(\mathbf{x}^*(t), \boldsymbol{\lambda}(t), u^*(t), t) \leq \mathcal{H}(\mathbf{x}^*(t), \boldsymbol{\lambda}(t), u(t), t) \quad (8)$$

Substituting the Hamiltonian from (6), Pontryagin's Minimum from (8) can be re-written.

$$1 + \lambda_x v \cos\theta^* + \lambda_\theta \frac{v}{R} u^* \leq 1 + \lambda_x v \cos\theta^* + \lambda_\theta \frac{v}{R} u \quad (9)$$

Simplifying the above expression in (9), the following relation is obtained, describing the optimal control.

$$\lambda_\theta(t) u^*(t) \leq \lambda_\theta(t) u(t) \Rightarrow u^*(t) = -\operatorname{sign}(\lambda_\theta(t)) \quad (10)$$

This optimal control in (10) hold the property required for the minimization of the Hamiltonian.

$$u^*(t) = \operatorname*{argmin}_{-1 \leq u(t) \leq 1} \{\mathcal{H}\} \Rightarrow u^*(t) = -\operatorname{sign}(\lambda_\theta(t)) \quad (11)$$

By the transversality condition, the terminal costates are obtained as a function of terminal state, control, and time [23].

$$\boldsymbol{\lambda}_f = \left.\frac{\partial \Phi(\cdot)}{\partial \mathbf{x}}\right|_{t=t_f} + \sigma \left.\frac{\partial \varphi(\cdot)}{\partial \mathbf{x}}\right|_{t=t_f} \quad (12)$$

In the transversality equation[1] in (12), the slack variable, $\sigma$, is introduced, the zero-level-set of the $\varphi(\cdot) = x$ gives the manifold $\mathcal{M}$, the terminal cost, $\Phi(\cdot)$, is zero as seen in the objective cost functional in (5). Evaluating (12), the terminal costate values are

$$\lambda_{x_f} = \sigma, \quad \lambda_{\theta_f} = 0. \quad (13)$$

The Hamiltonian at final time is zero [23].

$$\mathcal{H}^*(t_f) = -\left.\frac{\partial \Phi(\cdot)}{\partial t}\right|_{t=t_f} - \sigma \left.\frac{\partial \varphi(\cdot)}{\partial t}\right|_{t=t_f} = 0 \quad (14)$$

Substituting (11) and (13) back into (6) and evaluating the Hamiltonian at the final time,

$$\mathcal{H}^*(t_f) = 0 = 1 + \lambda_{x_f} v \cos\theta_f - \lambda_{\theta_f} \frac{v}{R} \operatorname{sign}(\lambda_{\theta_f}). \quad (15)$$

further algebraic manipulation of (15) and substituting the costate values at final time from (13), the optimal Hamiltonian allows one to solve for the slack variable, $\sigma$.

$$0 = 1 + \sigma v \cos\theta_f \Rightarrow \sigma = -\frac{1}{v \cos\theta_f} \quad (16)$$

From (16), the costates at final time (when the state is on the terminal manifold) are now known as a function of the final state and problem parameters:

---

[1]Note that the explicit dependence upon states, costates, control, and time have been dropped for reasons of space the notation of $(\cdot)$ is a surrogate for $(\mathbf{x}(t), \boldsymbol{\lambda}(t), u(t), t)$

$$\lambda_{x_f} = -\frac{1}{v \cos\theta_f}, \quad \lambda_{\theta_f} = 0. \quad (17)$$

Over the domain of the UP, $-\frac{\pi}{2} < \theta_f < \frac{\pi}{2}$, $\cos(\theta_f) > 0$, therefore $\lambda_{x_f} < 0$. Also, note that $\lambda_{x_f}$ is negative and $\lambda_{\theta_f} = 0$ as the costates at final time are aligned with the outward pointing normals.

The system dynamics under optimal control (11) are as follows.

$$\begin{aligned}
\dot{x}(t) &= v \cos(\theta(t)), & x(t=0) &= x_0, \\
\dot{\theta}(t) &= -\frac{v}{R}\operatorname{sign}(\lambda_\theta(t)), & \theta(t=0) &= \theta_0, \\
\dot{\lambda}_x(t) &= 0, & \lambda_x(t=0) &= \lambda_x, \\
\dot{\lambda}_\theta(t) &= \lambda_x v \sin(\theta(t)), & \lambda_\theta(t=0) &= \lambda_{\theta_0}
\end{aligned} \quad (18)$$

In retrograde time the dynamics are as follows.

$$\begin{aligned}
\mathring{x}(\tau) &= -v\cos(\theta(\tau)), & x(\tau=0) &= x_f = 0, \\
\mathring{\theta}(\tau) &= \frac{v}{R}\operatorname{sign}(\lambda_\theta(\tau)), & \theta(\tau=0) &= \theta_f, \\
\mathring{\lambda}_x(\tau) &= 0, & \lambda_x(\tau=0) &= \lambda_x, \\
\mathring{\lambda}_\theta(\tau) &= -\lambda_x v \sin(\theta(\tau)), & \lambda_\theta(\tau=0) &= \lambda_{\theta_f} = 0
\end{aligned} \quad (19)$$

The UP as described in (4) can be broken into three parts to assist explanation of optimal trajectories for reaching the UP in minimum time: (1) the case where $0 < \theta_f < \frac{\pi}{2}$, (2) the case where $-\frac{\pi}{2} < \theta_f < 0$, and (3) the case where $\theta_f = 0$. As will be seen, symmetry about $\theta_f = 0$ exists and the only difference is the turn-direction taken by the vehicle.

### A. Dispersal Line

**Lemma 1**. There exists a *dispersal line* (DL) in the state space given by the following:

$$\text{DL} = \{(x, \theta)|\ x < 0, \theta = \pi\}. \quad (20)$$

*Proof*: Assuming that the vehicle starts with a heading $\theta_0 = \pi$, the vehicle is pointing directly away from the terminal manifold. Any optimal trajectory beginning with $u^*(t=0) = 1$ can be reflected about the $x$-axis without effecting the time-to-go – the time to reach the terminal manifold. The time it takes to reach the manifold is equivalent if the vehicle turns counter clockwise $u(t) = 1$ or clockwise $u(t) = -1$ at initial time. This equivalence shows the existence of a DL. ∎

*Remark*. Trajectories cannot be back-propagated past the DL due to its singular nature. No Trajectories cross the DL.

*Remark*. The DL arrises due to symmetry and is comparable to other DLs in literature [28]–[30].

### B. Turn Only Strategy

**Lemma 2**. If $\theta_f \in \left(-\frac{\pi}{2}, \frac{\pi}{2}\right] \setminus 0$, then the optimal control is constant.

*Proof*: Suppose that the vehicle reaches the UP and has terminal angle $0 < \theta_f < \frac{\pi}{2}$. From (19), one observes that the costate

$\lambda_\theta(\tau)$ is increasing in retrograde time at $\tau = 0$; this is because $\mathring{\lambda}_\theta(\tau = 0) > 0$. In fact, so long as $\theta > 0$, $\mathring{\lambda}_\theta > 0$. This implies that $\lambda_\theta > 0$ if $\theta > 0$. From the optimal control (11), $u^*(t) = -1$ because $\lambda_\theta > 0$. Because $u^*(t) = -1$, we see from the retrograde equations (19) that $\theta$ is increasing. Therefore, $u^*(t)$ is constant over the trajectory when $0 < \theta < \pi$. By symmetry, when $-\frac{\pi}{2} < \theta_f < 0$, one can state that $u^*(t) = 1$ and remains constant when $-\pi < \theta < 0$. ∎

Now that the optimal control is obtained, one may integrate the retrograde equations in (19) to obtain the optimal trajectories that reach the UP with $\theta_f \ne 0$ at final time.

Integrating $\mathring{\theta}(\tau)$ in (19), when $0 < \theta_f < \frac{\pi}{2}$,

$$\mathring{\theta}(\tau) = \frac{v}{R}, \quad \theta(\tau = 0) = \theta_f \tag{21}$$
$$\Rightarrow \theta(\tau) = \tfrac{v}{R}\tau + \theta_f$$

Thus the final time is simply

$$t_f = \frac{R}{\nu}(\theta_0 - \theta_f). \tag{22}$$

Substitution of $\theta(\tau)$ into the retrograde equations (19), the retrograde equations for $x(\tau)$ can be solved.

$$\mathring{x}(\tau) = -v\cos\left(\tfrac{v}{R}\tau + \theta_f\right), \quad x(\tau=0) = x_f$$
$$\Rightarrow x(\tau) = -R\sin\left(\tfrac{v}{R}\tau + \theta_f\right) + R\sin(\theta_f) \tag{23}$$
$$\Rightarrow x(\theta) = -R(\sin(\theta) - \sin(\theta_f)), \quad \theta \in [\theta_f, \pi]$$

Therefore by symmetry, the state $x$ as a function of the vehicle heading $\theta$ is as follows.

$$x(\theta) = \begin{cases} R(\sin\theta - \sin\theta_f), & \theta_f \in (-\tfrac{\pi}{2}, 0) \wedge \theta \in (-\pi, \theta_f] \\ R(\sin\theta_f - \sin\theta), & \theta_f \in (0, \tfrac{\pi}{2}) \wedge \theta \in [\theta_f, \pi] \end{cases} \tag{24}$$

The optimal control when $\theta_f \ne 0$ is as follows.

$$u^*(t) = \begin{cases} -1, & \theta > 0 \\ +1, & \theta < 0 \end{cases} \tag{25}$$

The result in (25) stems from the proof of Lemma 2. Lastly, the final time, for a turn-only strategy is given by

$$t_f = \frac{R}{\nu}|\theta_0 - \theta_f|. \tag{26}$$

### C. Turn-Straight Strategy

Suppose that the vehicle reaches the UP and has terminal angle $\theta_f = 0$. In this case, the vehicle reaches the UP by way of reaching a *Universal Line* (UL). The strategy is one of turn-straight as the UL lies on the $x$-axis and represents straight-line trajectories.

From the retrograde dynamics in (19), if the terminal angle $\theta_f = 0$, then $\mathring{\lambda}_\theta(\tau = 0) = 0$. Because $\lambda_{\theta_f} = 0$, the value of $\lambda_\theta(\tau = 0^+) = 0$. From the Hamiltonian in (6), when $\lambda_\theta = 0$, the control vanishes and is undefined. In particular $u(\tau = 0^+) = $ undef – this suggests the presence of a singular surface spanning the $x$-axis, where $\theta = 0$.

**Lemma 3**. There exists a Universal Line given by the following.

$$\text{UL} = \{(x, \theta) \mid x < 0, \theta = 0\}. \tag{27}$$

The control, $u^*(t) = 0$, on the UL and remains zero until the state terminates on the UP.

*Proof*: The shortest path between a point (i.e, the location of the vehicle) and a line is along a straight line that is perpendicular to the the former. When $u = 0$, the vehicle is moving on a straight line, and and when $\theta = 0$, the vehicle is heading directly towards the line. Thus it's heading is perpendicular to the line. Moreover, this singular control satisfies the Hamilton-Jacobi-Bellman (HJB) equation as shown in the following.

$$\frac{\partial V}{\partial t} + \min_u (\nabla_\mathbf{x} V \cdot \mathbf{f}) = 0. \tag{28}$$

Here, $V := \min_u J$ is the Value function, and, by definition, $\nabla_\mathbf{x} V = \boldsymbol{\lambda}$. The time-to-go under the prescribed control is simply the distance over the velocity: $V = -\frac{x}{v}$. Substituting the control $u = 0$, along with $\theta = 0$, into the HJB gives the following.

$$0 = 1 + \min_u \left( \begin{bmatrix} -\tfrac{1}{v}, & 0 \end{bmatrix}^\top \cdot \begin{bmatrix} v\cos\theta, & \tfrac{v}{R}u \end{bmatrix}^\top \right) \tag{29}$$
$$0 = 1 - 1$$
∎

Consider the tributaries of the UL. For these trajectories, the UL may be considered to be the terminal surface wherefrom the optimal time-to-go to the line is known. Let $t_s$ be the time at which the vehicle reaches the UL ("s" for "switch"). Thus the associated function whose zero level-set is the UL, and the remaining time-to-go are given by the following.

$$\varphi^{\text{UL}}(x, \theta) = \theta, \quad \Phi^{\text{UL}} = -\frac{x_S}{v} \tag{30}$$

The transversality conditions yields the following terminal costates.

$$\boldsymbol{\lambda}^{\text{UL}}(t_s) = \frac{\partial \Phi^{\text{UL}}}{\partial \mathbf{x}_s} + \sigma^{\text{UL}} \frac{\partial \varphi^{\text{UL}}}{\partial \mathbf{x}_s}$$
$$= \begin{bmatrix} -\tfrac{1}{v}, & 0 \end{bmatrix}^\top + \sigma^{\text{UL}} [0, 1]^\top \tag{31}$$
$$\Rightarrow \lambda_{x_s} = -\frac{1}{v}, \quad \lambda_{\theta_s} = \sigma^{\text{UL}}$$

Thus, $\lambda_{x_s} < 0$ as before. At $t = t_s$, $\theta$ goes to zero; thus the tributary must have come from $\theta \ne 0$ at a time just before $t_s$, i.e., $t_{s^-}$. Therefore, the control, $u_{s^-}$, must be nonzero as well. Moreover, at $t_{s^-}$, the optimal control, (11), applies and thus $u^*_{s^-} \in \{-1, 1\}$. When $\theta_{s^-} > 0$ (resp. $< 0$) it must be the case that $u^*_{s^-} = -1$ (resp. $= 1$) in order to drive $\theta$ to zero at $t_s$. This implies, from (11), that $\lambda_{\theta_{s^-}} > 0$ (resp. $< 0$). From the same logical arguments described in the proof of Lemma 2, $u^*$ must be constant (either $-1$ or $1$) up until time $t_s$. Integration of the retrograde dynamics, (19), from a point $(x_s, 0)$ on the UL yields the following flowfield equations.

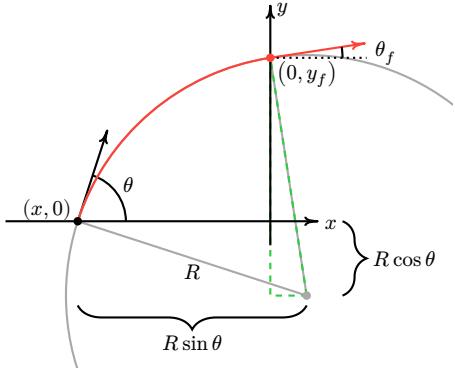

Fig. 3. Geometry for turn only trajectories

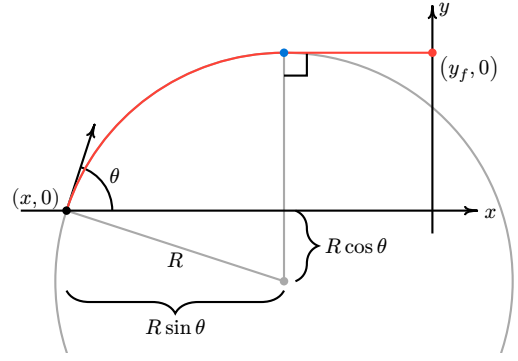

Fig. 4. Geometry for turn-straight trajectories

$$x(\theta) = \begin{cases} R\sin\theta + s, & \theta \in (-\pi, 0] \\ -R\sin\theta + s, & \theta \in [0, \pi] \end{cases} \quad (32)$$

The optimal control for turn-straight is as follows.

$$u^*(t) = \begin{cases} -1, & \theta > 0 \\ 0, & \theta = 0 \\ +1, & \theta < 0 \end{cases} \quad (33)$$

When the vehicle is on the UL, the strategy is a straight only trajectory - the turn time is zero.

$$u^S(t) = 0, t \in [0, t_f]. \quad (34)$$

## III. GEOMETRIC CONSTRUCTION

With the optimal control and flowfield fully characterized, the geometry in the Cartesian coordinate system may be used to obtain the escape time and location. Consider the position of the vehicle as $(x, y)$ coordinates with a particular heading, $\theta$, measured w.r.t. the positive $x$-axis. Without loss of generality, let the vehicle's initial $y$ location be set to 0. The geometry for the turn-only and turn-straight trajectories are shown in Fig. 3 and Fig. 4, respectively.

The first step is to determine, based on the vehicle's state, $(x, y, \theta)$, whether the optimal trajectory is turn only or turn-straight. From the results in Section II.B, the most limiting case for turn only is when $\theta_f = 0$. The entire trajectory, obtained by substituting $\theta_f = 0$ into (24), therefore creates a partition in the state space, as follows.

$$\begin{aligned} \mathcal{R}^T &= \{(x, \theta) \mid x \geq -R\sin|\theta|, \theta \in (-\pi, \pi]\} \\ \mathcal{R}^{TS} &= \{(x, \theta) \mid x < -R\sin|\theta|, \theta \in (-\pi, \pi]\} \end{aligned} \quad (35)$$

The subscripts $T$ and $TS$ denote "turn only" and "turn-straight".

Consider the turn only geometry in Fig. 3. The quantity $y_f$ represents the escape location along the line. To obtain $y_f$, it is useful to consider the green dashed right triangle. The horizontal side is $R\sin|\theta| + x$ (recalling that $x < 0$); the vertical side is $y_f + R\cos\theta$; finally, its hypotenuse is simply $R$. Thus the following expression yields $y_f$.

$$y_f^T = \text{sign}(\theta)\left[\sqrt{R^2 - (R\sin|\theta| + x)^2} - R\cos\theta\right] \quad (36)$$

The same triangle can be used to obtain the final heading.

$$\theta_f = \text{sign}(\theta)\sin^{-1}\left(\frac{R\sin|\theta| + x}{R}\right) \quad (37)$$

Finally, substituting the final angle into (26) yields the following expression for the escape time.

$$t_f^T = \frac{R}{v}||\theta| - |\theta_f|| \quad (38)$$

The associated optimal control is given by the following.

$$u^T(t) = -\text{sign}(\theta), t \in [0, t_f] \quad (39)$$

Consider the turn-straight geometry in Fig. 4. The escape location is given by the following.

$$y_f^{TS} = \text{sign}(\theta)R(1 - \cos\theta) \quad (40)$$

The escape time is simply the time required to drive $\theta$ to zero plus the time spent to traverse the remaining straight-line segment.

$$t_f^{TS} = \tfrac{1}{v}(R|\theta| - x - R\sin|\theta|) \quad (41)$$

The associated optimal control is given by the following.

$$u^{TS}(t) = \begin{cases} -\text{sign}(\theta), & t \in \left[0, |\theta|\frac{R}{v}\right] \\ 0, & t \in \left(|\theta|\frac{R}{v}, t_f\right] \end{cases} \quad (42)$$

## IV. POLYGON ESCAPE

Leveraging the min-time solution for reaching an infinite line, the approach is extended to convex polygons. This approach requires that the unique, time-optimal solution for the vehicle to reach the line be calculated. The resulting time to escape and strategy can readily be obtained for each line of the polygon. Two checks are made:
1) Does the optimal trajectory entail a straight, turn-only, or turn-straight strategy?

**Algorithm 1:** Min-Time Convex Polygon Escape

1  Initiate `location`, `speed`, and `heading` of vehicle
2  Initiate the `verticies` of the polygon
3  **for** each edge, $i$, of the polygon
4     $(x, \theta) \leftarrow$ `LocalCoords(location, heading)`
5     **if** $\theta = 0$                                          # Straight-Only
6        $t_{f_i} \leftarrow -\frac{x}{v}$
7        $u_i \leftarrow u^S$, (34)
8     **else if** $(x, \theta) \in \mathcal{R}^T$                    # Turn-Only
9        $t_{f_i} \leftarrow$ (38)
10       $u_i \leftarrow u^T$, (39)
11    **else**                                               # Turn-Straight
12       $t_{f_i} \leftarrow$ (41)
13       $u_i \leftarrow u^{TS}$, (42)
14    **end**
15 **end**
16 $i^* \leftarrow \arg\min_i t_{f_i}$
17 **return** $t_f[i^*], u[i^*]$

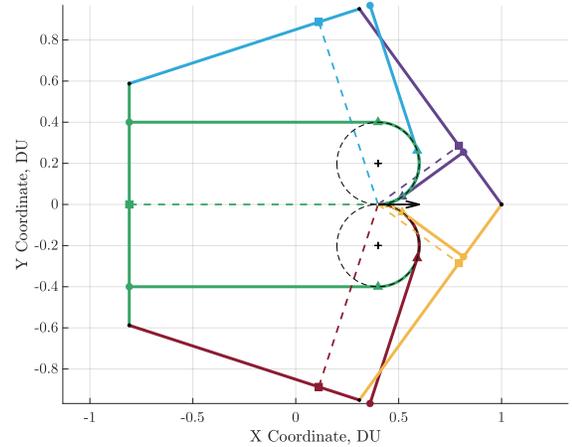

Fig. 6. An illustration of time-optimal equivalent strategies for escaping a polygon. The green trajectories result from the vehicle starting on a DL for the green edge of the polygon. The vehicle can also reach the purple and yellow edges with equivalent time. Because the vehicle departs the polygon prior to reaching the blue and red edges, these time-equivalent trajectories are of no consequence.

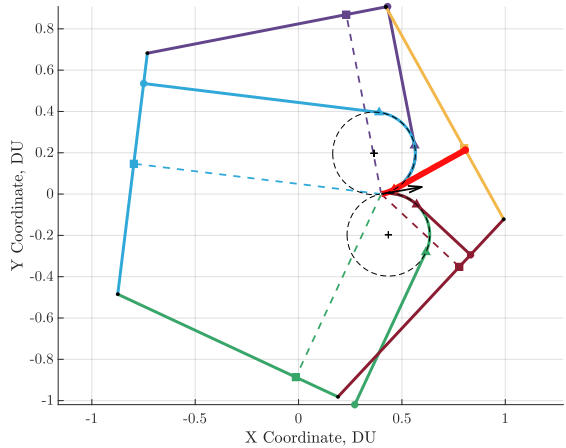

Fig. 5. The minimum time escape of a convex polygon is demonstrated. Highlighted in red is the min-time path to escape the polygon.

2) Do turning strategies require a right- or left-hand turn?

From these checks, one may compute all possible min-time strategies for each line segment of the polygon using the approach specified earlier and then select the strategy that reaches the line in min-time. Algorithm 1 summarizes the proposed procedure, and an example is shown in Fig. 5.

*Remark.* It is possible that two such trajectories are equally optimal, i.e., there may exist dispersal *surface(s)*. These could arise, for example, when the vehicle is pointed directly away from the edge providing the min time escape, or when the vehicle is positioned on and pointing along the angular bisector of two min time edges.

A figure that illustrates these two kinds of Dispersal Lines and equivalent trajectories is seen in Fig. 6.

## V. CONCLUSION

In conclusion, a turn constrained vehicle striving to escape a polygon in minimum time was considered. Using the method of characteristics, the min-time strategies for reaching a line of infinite length were first solved. The resulting time-optimal strategies were then used to identify critical portions of the terminal manifold and structures within the state space including Usable Parts, a Dispersal Line, and a Universal Line. Next, the methods for reaching the line were extended for escaping a polygon. Using this constructive technique, the min-time path to reach each edge was obtained. The resulting minimum of the set of optimal trajectories is then selected for escaping the polygon. A demonstration of how this works was also communicated in an Algorithm. Future work entails extending this work to non-convex polygons as well as more general geometries including those constructed of arcs, lines, and splines.

## REFERENCES

[1] L. E. Dubins, "On Curves of Minimal Length with a Constraint on Average Curvature, and with Prescribed Initial and Terminal Positions and Tangents," *American Journal of Mathematics*, vol. 79, no. 3, pp. 497–516, 1957.
[2] L. E. Dubins, "On Plane Curves with Curvature," *Pacific Journal of Mathematics*, vol. 11, no. 2, pp. 471–481, 1961.

[3] J. Reeds and L. Shepp, "Optimal paths for a car that goes both forwards and backwards," *Pacific journal of mathematics*, vol. 145, no. 2, pp. 367–393, 1990.

[4] L. S. Pontryagin, V. G. Boltyanskii, R. V. Gamkrelidze, and E. F. Mishchenko, *The Mathematical Theory of Optimal Processes*. New York–London: Interscience Publishers John Wiley & Sons, Inc., 1962.

[5] J.-D. Boissonnat, A. Cérézo, and J. Leblond, "Shortest paths of bounded curvature in the plane," *Journal of Intelligent and Robotic Systems*, vol. 11, pp. 5–20, 1994.

[6] A. M. Shkel and V. Lumelsky, "Classification of the Dubins set," *Robotics and Autonomous Systems*, vol. 34, no. 4, pp. 179–202, 2001, doi: 10.1016/S0921-8890(00)00127-5.

[7] J.-D. Boissonnat and S. Lazard, "A polynomial-time algorithm for computing a shortest path of bounded curvature amidst moderate obstacles," in *Proceedings of the twelfth annual symposium on Computational geometry*, 1996, pp. 242–251. doi: 10.1145/237218.237393.

[8] P. K. Agarwal, P. Raghavan, and H. Tamaki, "Motion planning for a steering-constrained robot through moderate obstacles," in *Proceedings of the twenty-seventh annual ACM symposium on Theory of Computing*, 1995, pp. 343–352. doi: 10.1145/225058.225158.

[9] T. G. McGee and J. K. Hedrick, "Optimal path planning with a kinematic airplane model," *Journal of guidance, control, and dynamics*, vol. 30, no. 2, pp. 629–633, 2007, doi: 10.2514/1.25042.

[10] E. Bakolas and P. Tsiotras, "Optimal synthesis of the Zermelo–Markov–Dubins problem in a constant drift field," *Journal of Optimization Theory and Applications*, vol. 156, no. 2, pp. 469–492, 2013.

[11] L. Techy and C. A. Woolsey, "Minimum-time path planning for unmanned aerial vehicles in steady uniform winds," *Journal of guidance, control, and dynamics*, vol. 32, no. 6, pp. 1736–1746, 2009.

[12] K. Mittal, J. Song, S. Gupta, and T. A. Wettergren, "Rapid path planning for Dubins vehicles under environmental currents," *Robotics and Autonomous Systems*, vol. 134, p. 103646–103647, 2020.

[13] A. Wolek and C. Woolsey, "Feasible Dubins paths in presence of unknown, unsteady velocity disturbances," *Journal of Guidance, Control, and Dynamics*, vol. 38, no. 4, pp. 782–787, 2015.

[14] A. Wolek, "Optimal paths in gliding flight," *PhD Thesis, Virginia Tech, Blacksburg, VA*, 2015.

[15] Z. Chen and T. Shima, "Shortest Dubins paths through three points," *Automatica*, vol. 105, pp. 368–375, 2019.

[16] X. Ma and D. A. Castanon, "Receding horizon planning for Dubins traveling salesman problems," in *Proceedings of the 45th IEEE Conference on Decision and Control*, 2006, pp. 5453–5458.

[17] K. Savla, E. Frazzoli, and F. Bullo, "Traveling Salesperson Problems for the Dubins Vehicle," *IEEE Transactions on Automatic Control*, vol. 53, no. 6, pp. 1378–1391, 2008, doi: 10.1109/TAC.2008.925814.

[18] P. Váňa and J. Faigl, "On the dubins traveling salesman problem with neighborhoods," in *2015 IEEE/RSJ International Conference on Intelligent Robots and Systems (IROS)*, 2015, pp. 4029–4034.

[19] A. Wolek, J. McMahon, B. R. Dzikowicz, and B. H. Houston, "The Orbiting Dubins Traveling Salesman Problem: planning inspection tours for a minehunting AUV," *Autonomous Robots*, vol. 45, no. 1, pp. 31–49, 2021.

[20] T. L. Molloy and I. Shames, "Minimum-time escape from a circular region for a Dubins car," *IFAC-PapersOnLine*, vol. 56, no. 1, pp. 43–48, 2023.

[21] I. Weintraub, A. Von Moll, and M. Pachter, "Minimum Time Escape from a Circular Region of a Dubins Car," in *IEEE National Aerospace and Electronics Conference*, 2024.

[22] R. Isaacs, *Differential Games: A Mathematical Theory with Applications to Warfare and Pursuit, Control and Optimization*. John Wiley, Sons, 1965.

[23] A. E. Bryson, *Applied optimal control: optimization, estimation and control*. Routledge, 2018.

[24] D. E. Kirk, *Optimal Control Theory: An Introduction*. Dover, 2004.

[25] S. G. Manyam, A. Nayak, and S. Rathinam, "G*: A New Approach to Bounding Curvature Constrained Shortest Paths through Dubins Gates.," in *Robotics: Science and Systems*, 2023.

[26] P. Váňa and J. Faigl, "On the Dubins traveling salesman problem with neighborhoods," in *2015 IEEE/RSJ International Conference on Intelligent Robots and Systems (IROS)*, 2015, pp. 4029–4034.

[27] S. G. Manyam and S. Rathinam, "On tightly bounding the Dubins traveling salesman's optimum," *Journal of Dynamic Systems, Measurement and Control*, vol. 140, no. 7, p. 71013–71014, 2018.

[28] Z. Akilan and Z. Fuchs, "Zero-sum turret defense differential game with singular surfaces," in *2017 IEEE Conference on Control Technology and Applications (CCTA)*, 2017, pp. 2041–2048.

[29] A. W. Merz, "The homicidal chauffeur," *AIAA Journal*, vol. 12, no. 3, pp. 259–260, 1974.

[30] T. Başar and G. J. Olsder, *Dynamic noncooperative game theory*. SIAM, 1998.